\documentclass[12pt]{iopart}
\usepackage{amssymb}

\newcommand{\cA}{{\cal A}}

\newcommand{\cC}{{\cal C}}
\newcommand{\cD}{{\cal D}}
\newcommand{\cE}{{\cal E}}

\newcommand{\cH}{{\cal H}}
\newcommand{\cJ}{{\cal J}}

\newcommand{\cO}{{\cal O}}
\newcommand{\cP}{{\cal P}}
\newcommand{\cQ}{{\cal Q}}
\newcommand{\cR}{{\cal R}}
\newcommand{\cS}{{\cal S}}
\newcommand{\cU}{{\cal U}}
\newcommand{\cV}{{\cal V}}

\newcommand{\cmplxs}{{\mathbb C}}
\newcommand\openone{\leavevmode\hbox{\small1\normalsize\kern-.33em1}}

\begin{document}

\title{Constructing Qubits in Physical Systems}
\author{Lorenza Viola,  Emanuel Knill, and Raymond Laflamme}
\address{
Los Alamos National Laboratory, MS B265, Los Alamos, 
New Mexico 87545, \vspace*{2.5mm}USA
E-mail: {\tt lviola@lanl.gov, knill@lanl.gov, laflamme@lanl.gov } }

\begin{abstract}
The notion of a qubit is ubiquitous in quantum information processing.
In spite of the simple abstract definition of qubits as two-state
quantum systems, identifying qubits in physical systems is often
unexpectedly difficult.  There are an astonishing variety of ways in
which qubits can emerge from devices.  What essential features are
required for an implementation to properly instantiate a qubit?  We
give three typical examples and propose an operational
characterization of qubits based on quantum observables and
subsystems.
\end{abstract}

\pacs{03.67.-a,03.67.Lx,89.70.+c}
\ams{81P68,68Q05,15A69}

\maketitle

\section{Introduction}

Quantum bits ({\sl qubits}) are the elementary units of information
that are used to represent quantum data~\cite{schumacher:1996}. Thus,
the idea of a qubit underlies all investigations in the rapidly
growing science of quantum information -- including quantum
information theory, quantum communication, quantum computation,
quantum complexity, and quantum game
theory~\cite{bennett:2000,preskill,feynman,nielsen}.  In particular,
qubits are the basic building blocks for defining the standard model
of quantum computation as introduced by Deutsch~\cite{deutsch}, which
has so far provided the appropriate representation for identifying and
understanding {\sl efficient} ways of processing information using
quantum mechanics. Its investigation resulted in feasible algorithms
for factoring large integers~\cite{shor} and for simulating
many-particle quantum systems~\cite{lloyd:1996}, two problems not
known to be efficiently solvable with classical computers.

A qubit can be thought of as the extension of a classical bit obtained
by applying the superposition principle. When quantum superposition 
states of many qubits are constructed in the tensor product state space 
that quantum mechanics prescribes for a composite system, quantum 
entanglement arises as an additional information resource with no 
classical counterpart. However, qubits share with classical
bits the fundamental property of being a {\sl fungible} information
resource~\cite{bennett,toffoli}: While both classical and quantum
information is intended, in fact required, to be physically realized,
it is {\sl abstractly} defined and therefore independent of the
details of the underlying physical realization.

The fungibility property is essential to quantum information in two
respects. First, by defining quantum information independent of the
details of specific physical devices and their complex physics, it has
been possible to study qubit properties at the abstract level and thus
to obtain a deeper understanding of the distinctive features that
qubits inherit from their intrinsic quantum-mechanical nature.
Examples of fundamental results following from the basic properties of
superpositions and of the randomness associated with quantum
measurements include the fact that qubits in an unknown quantum state
cannot be perfectly copied (no-cloning theorem \cite{wooters,dieks})
and they cannot be broadcast (no-broadcasting theorem
\cite{barnum}). On the other hand they can be reliably communicated by
means of the quantum teleportation protocol~\cite{bennett:1993}, an
extremely useful protocol with many applications~\cite{gottesman}.
Second, from a practical standpoint, the arbitrariness of the physical
realization implied by fungibility allows for greater flexibility in
the identification and design of quantum information processors. This
is reflected in the amazing variety of representations of qubits that
have appeared in recent proposals for physical realizations of quantum
computers~\cite{expprop}. While such variety greatly increases the
number of physical systems with the potential for quantum information
processing, it can be intimidating when considering a new prospective
system for implementing qubits.  As recently observed by
DiVincenzo~\cite{divince:2000}, ``recognizing a qubit can be trickier
than one might think''.  Thus to determine the suitability of a
physical system for quantum information processing we need to answer
the following basic questions: What is a qubit, and how do we look for
them?

In this paper, we identify essential criteria to be met by a system to
contain physical embodiments of qubits. Our primary goal is to develop
the intuition needed for ``distilling'' qubits from physical systems,
in the hope that such intuition will serve as a guide for exploring
the full range of possibilities available for physical implementation.
We do so by revisiting a few prototypical examples to obtain a general
operational characterization of qubits.  The examples are examined in
detail in Section~\ref{sect:fungibility}.  The first example shows how
the state space of a collection of bosonic modes can be exploited for
representing qubits based on the requirement of obtaining realizable
control over the qubit observables.  The second and third example are
essentially motivated by reconsidering the notion of what a qubit is
in the light of our knowledge from quantum error-correcting
methods. In particular, we analyze two situations where a qubit can be
embedded into the larger state space of three physical two-state
systems in a way which makes it robust against noise to first-order in
time and to arbitrary orders in time, respectively.  As the level of
abstraction required for constructing our example qubits grows, the
qubit states are less easily related to states of the underlying
physical degrees of freedom, and more powerful mathematical tools are
required to describe the qubits. The emerging qubit criteria are
summarized in Section~\ref{sect:criteria}. Our analysis points to the
crucial role played by {\sl control resources} in determining a
preferred qubit realization, and to an operator picture in terms of
{\sl quantum observables}, rather than states, as the appropriate
framework for capturing the idea of a qubit in full generality.

\section{The fungibility of qubits}
\label{sect:fungibility}

The first proposals for quantum computers implemented qubits literally
by exploiting explicit two level systems.  Examples of qubits in these
proposals include the two levels corresponding to absence and
presence of a photon in a mode~\cite{milburn:qc1989}, spin $1/2$
nuclear spins \cite{lloyd:qc1993}, and the ground and one of the excited
states of ions \cite{cirac:qc1995a}.  The proposals have in common the
goal of implementing qubits at the fundamental level of a physical
system. The recognition of how fragile these qubits' states are led
to the development of quantum error correction, which necessarily
involves maintaining the information by distributing it in a
controlled way over a potentially large number of physical systems.
As a result we reconsidered the notion of a qubit.

To illustrate the different facets of the notion of a qubit we determine 
how an entity equivalent to an abstractly defined qubit can be constructed 
in three prototypical situations.  An abstractly defined qubit can be 
identified with an ideal two-state quantum system {\sf Q} whose
associated state space is a two-dimensional complex Hilbert space
$\cH_{\sf q} \simeq {\cmplxs}^2$.  An orthonormal basis $\{
|0\rangle_{\sf q}, |1\rangle_{\sf q} \}$ (computational basis) is
fixed in $\cH_{\sf q}$ and we make the identification $|0\rangle_{\sf
q} \simeq (1,0)$, $|1\rangle_{\sf q} \simeq (0,1)$ with vectors in
${\cmplxs}^2$.  Operations on states are conveniently expressed in
terms of the Pauli operators $\sigma_{\sf x},\sigma_{\sf y},
\sigma_{\sf z}$, which obey commutation and anti-commutation 
relationships of the form
\begin{eqnarray}
  [ \sigma_\alpha, \sigma_\beta ] & = &2i \,\sum_\gamma 
   \varepsilon_{\alpha\beta\gamma}
   \sigma_\gamma \:, \label{comm} \\
 \{ \sigma_\alpha, \sigma_\beta \} & = & 2 \,\delta_{\alpha\beta} \openone \:, 
\label{anticomm}
\end{eqnarray}
with $\alpha,\beta,\gamma \in \{ {\sf x,y,z} \}$, and
$\varepsilon_{\alpha\beta\gamma}$ and $\openone$ being the completely
antisymmetric symbol and the identity on $\cH_{\sf q}$,
respectively. To simplify notations, we define $X:= \sigma_{\sf x},
Y:= \sigma_{\sf y}, Z:= \sigma_{\sf z}$.

As we will see, the abstract state space of a qubit is not necessarily
directly related to the state space of the physical systems.  However,
the operators are naturally embedded in the operator algebras
associated with the physical observables.

\subsection{The bosonic qubit}

Bosonic qubits are the basic building blocks in a recent proposal for
linear optics quantum computation ({\sc loqc}) \cite{klm:2000}. The
relevant physical system is a collection of $2n$ distinguishable modes
which describe the elementary excitations of the quantized
electro-magnetic field \cite{walls}. Each mode is characterized by
annihilation and creation operators ${\bf a}_k, {\bf a}_k^\dagger$,
$k=1,\ldots, 2n$, which satisfy bosonic commutation rules $[ {\bf
a}_k, {\bf a}_{k'}]= [ {\bf a}_k^\dagger, {\bf a}_{k'}^\dagger]=0$, $[
{\bf a}_k^\dagger, {\bf a}_{k'}]=\delta_{k k'}$. In particular, number
states of mode $k$ are defined as eigenstates of the number operator
${\bf n}_k={\bf a}_k^\dagger {\bf a}_k$ for mode $k$, ${\bf n}_k
|n_k\rangle_k = n_k |n_k\rangle_k$. Number states constitute a basis
for the state space of a given mode {\it i.e.}, $\cH_k=\mbox{span}\{
|n_k\rangle_k \: | \: n_k=0,1,2,\ldots \}$ (Fock representation).
Accordingly,
\begin{equation}
\cS \simeq \cH_1 \otimes \cH_2 \ldots \otimes \cH_{2n}
\label{ss0}
\end{equation}
is the overall state space in which we have to look for qubits.  We
use the abbreviation $|n_1\rangle_1 \ldots |n_{2n}\rangle_{2n}
=|n_1 \ldots n_{2n}\rangle$ for product number states in $\cS$.

Clearly, a variety of (inequivalent) prescriptions are conceivable for
representing qubits in $\cS$. In {\sc loqc}, an encoding based on two
modes and one boson is adopted via the mapping
\begin{eqnarray}
|0\rangle_{{\sf q}(k,k')} & \rightarrow & |0\rangle_k \,|1\rangle_{k'} \:,
\nonumber \\
|1\rangle_{{\sf q}(k,k')} & \rightarrow & |1\rangle_k \,|0\rangle_{k'} \:,
\label{bq}
\end{eqnarray}
for a qubit supported by modes $(k,k')$. This choice is motivated by
the potential for easy implementations of single qubit transformations
via passive linear optical elements. It is worth pointing out that
encodings similar to the above, where the degree of freedom carrying
quantum information is identified with the presence of a
quasi-particle in one of two modes or sites, or the presence/absence
of a quasi-particle in a single mode or site, can make sense for
quantum statistics other than the bosonic one.  Thus, the present
discussion is relevant to situations where information is stored in
anyonic \cite{kitaev,lloyd:2000} or fermionic \cite{lloyd:1998,bravyi}
qubits. A different scheme for embedding a generic finite-dimensional
quantum system (or {\sl qudit}) in the state space of bosonic degrees
of freedom was recently proposed in~\cite{gottesman:2000}.

According to (\ref{bq}), the state space of the bosonic qubit, 
$\cH_{{\sf q}(k,k')}=\mbox{span}\{ |0\rangle_{{\sf q}(k,k')},  
|1\rangle_{{\sf q}(k,k')} \}$,
is identified with the one-excitation sector of the two-mode Hilbert space
$\cH_{(k,k')}=\cH_k \otimes \cH_{k'}$. By letting $n_{k, k'}$ denote the
eigenvalue of the joint number operator ${\bf n}_{k,k'}={\bf n}_{k}+
{\bf n}_{k'}$, one has formally
\begin{equation}
\cH_{(k,k')} = 
\cH_k \otimes \cH_{k'} \simeq \bigoplus_{n_{k,k'}=0}^\infty
\cH^{(n_{k,k'})} \simeq \cH_{{\sf q}(k,k')} \oplus \tilde{\cH}_k\:,
\label{sqss}
\end{equation}
where $\cH^{(n_{k,k'})}$ is the subspace of states in $\cH_{(k,k')}$ with 
$n_{k,k'}$ bosons, and 
$\tilde{\cH}_k =\oplus_{\{ n_{k,k'}\neq 1 \}} \cH^{(n_{k,k'})}$. 
The state space of $n$ bosonic qubits, which is obtained via an effective
tensor-product construction of one-qubit spaces, likewise relates to the 
overall state space $\cS$ in a non-trivial way. Assuming a pairing between
modes of the form $(k,k')=(k,k+1)$ for the $i$'th encoded qubit, $i=(k+1)/2
=1,\ldots,n$, we obtain
\begin{eqnarray}
\fl (\cH_1 \otimes \cH_2) \otimes \ldots \otimes (\cH_{2n-1} \otimes \cH_{2n}) 
 = 
\cH_{(1,2)} \otimes \cH_{(3,4)} \otimes \ldots \otimes \cH_{(2n-1,2n)} 
\nonumber \\ \lo \mbox{} \simeq 
\bigotimes_{  k=1,3,\ldots, 2n-1 } 
\Big[ \bigoplus_{n_{k,k+1}=0}^\infty \cH^{(n_{k,k+1})} \Big] 
\simeq \Big[ \cH_{{\sf q}_1} \otimes \cH_{{\sf q}_2} \ldots \otimes 
\cH_{{\sf q}_n} \Big] \oplus \tilde{\cH}\:,
\label{mqss}
\end{eqnarray}
where for each two-mode state space $\cH_{(k,k+1)}$ a decomposition similar
to (\ref{sqss}) has been used, $\cH_{{\sf q}_i}=\cH_{{\sf q}(2i-1,2i)}$, 
and $\tilde{\cH}$ collects all the contributions with excitation different 
from one in at least one of the pairs of modes.

Several remarks on the above state space structure are in order. In
the one-qubit case, the presence of the summand $\tilde{\cH}_k$ in
(\ref{sqss}) causes the bosonic qubit to be effectively embedded into a 
larger state space where infinitely many ``non-qubit'' levels are present.
One can formally rewrite this separation so that the additional levels
form part of an effective environment~\cite{tian:2000}.  
The formal state space of the two resulting interacting subsystems is 
written as 
\begin{equation}
\cH_{\sf qe}=(\cH_{{\sf q}(k,k')}\oplus|{v_{\sf q}}\rangle)
\otimes(\tilde{\cH}_k\oplus|{v_{\sf e}}\rangle) \:, 
\label{subsys}
\end{equation} 
where $|{v_{\sf q}}\rangle$ and $|{v_{\sf e}}\rangle$ are adjoined 
``vacuum'' states, and we make the identification 
$\cH_{{\sf q}(k,k')} \rightarrow \cH_{{\sf q}(k,k')} 
\otimes|{v_{\sf e}}\rangle$.  The idea generalizes to $n$ qubits, in 
which case the role of the effective ``non-qubit'' environment 
is assumed by $\tilde{\cH}$ in (\ref{mqss}).

Whenever a non-zero amplitude is found in the space $\tilde{\cH}$, one
or more qubits have ``leaked-out'' from the intended logical space
$\cH_L=\otimes_i \cH_{{\sf q}_i}$ so that the state no longer maps to
a state of qubits.  To guarantee that the state belongs to the qubits
for {\sl all} times $t$ means restricting the overall evolution
operator $\cU(t)$ to the sub-manifold of unitary operators $U(\cH_L)
\oplus U(\tilde{\cH}) \subset U(\cS)$. It is essential to realize that
such a requirement is in general unnecessary as well as overly
restrictive. While care should be taken to ensure that no leakage
allows the qubits to stray away from the logical space at the
beginning and at the end of every quantum gate, one should not insist
on retaining a mapping into qubits at every intermediate time.  In
other words, qubits need only exist {\sl stroboscopically} in time.

In the {\sc loqc} proposal, relaxing the constraint of well-defined 
qubit states at intermediate steps is indispensable for achieving the 
required two-bit conditional dynamics.
For example, in~\cite{klm:2000}, the most basic conditional sign-flip gate 
{\sf c-}$\sigma_{\sf z}$ on two bosonic qubits, say ${\sf q}_1={\sf q}(1,2)$
and ${\sf q}_2={\sf q}(3,4)$, is decomposed into a sequence of operations
\begin{equation}
\mbox{\sf c-}\sigma_{\sf z}^{({\sf q}_1,{\sf q}_2)} =
U^{(1,3)\dagger}_{\mbox{\sc bs}}
U_{\mbox{\sc ns}_1}U_{\mbox{\sc ns}_3}U^{(1,3)}_{\mbox{\sc bs}} \in 
U(\cH_{{\sf q}_1} \otimes \cH_{{\sf q}_2}) \oplus U(\tilde{\cH})\:, 
\label{csign}
\end{equation}
where both the beam-splitter gate $U^{(1,3)}_{\mbox{\sc bs}}$ and the
non-deterministic gate $U_{\mbox{\sc ns}_j}$ on mode $j=1,3$ cause a
temporary departure from the one-excitation space of each qubit. This
implementation is also non-deterministic, meaning that it requires
post-selection, with a known success probability. In~\cite{klm:2000},
the non-determinism can be removed by exploiting the quantum
teleportation protocol \cite{bennett:1993,gottesman}. 

In spite of the complicated underlying state space structure, a bosonic 
qubit ${\sf q}(k,k')$ can be straightforwardly characterized in terms of 
its generating observables. On the $2n$-mode state space $\cS$, define 
the operator
\begin{equation}
P_{k,k'}:= {1 \over 2\pi} \int_0^{2\pi} 
d\varphi\, e^{-i({\bf n}_k + {\bf n}_{k'}-1)\varphi}
\label{bpo}
\end{equation}
on $\cH_{(k,k')}$, and $\openone$ elsewhere. $P_{k,k'}$ satisfies
$P_{k,k'}=P_{k,k'}^\dagger =P_{k,k'}^2$, and 
\begin{equation}
P_{k,k'} |n_1 n_2 \ldots n_{2n} \rangle = \delta_{n_k + n_{k'},1}\,
         |n_1 n_2 \ldots n_{2n} \rangle\:. 
\label{bpo2}
\end{equation}
Hence, the restriction $\hat{P}_{k,k'}$ of $P_{k,k'}$ to $\cH_{(k,k')}$, 
$\hat{P}_{k,k'}: \cH_{(k,k')} \rightarrow \cH_{{\sf q}(k, k')}$,
and the product $\cP$ := $P_{1,2}P_{3,4}\ldots P_{2n-1,2n}$,
$\cP: \cS \rightarrow \cH_{{\sf q}_1}\otimes \ldots \otimes 
\cH_{{\sf q}_n}$, are the projectors onto the one-qubit and the 
$n$-qubit space, respectively. 
Clearly, the action of $\hat{P}_{k,k'}$ is the identity operation when 
further restricted to the qubit space, 
{\it i.e.} $\hat{P}_{k,k'}=_{\sf q}\openone_{{\sf q}(k,k')}$,
where $=_{\sf q}$ means equality over $\cH_{{\sf q}(k,k')}$.
For fixed $k,k'$, one finds that
\begin{equation}
\hat{P}_{k,k'}({\bf n}_k + {\bf n}_{k'})\hat{P}_{k,k'}=
\hat{P}_{k,k'}({\bf n}_k + {\bf n}_{k'})=
({\bf n}_k + {\bf n}_{k'})\hat{P}_{k,k'}\:.
\label{bid}
\end{equation}
From this one can readily check that the following operators act as encoded
generating observables for the bosonic qubit ${\sf q}(k,k')$, with $k'=k+1$:
\begin{eqnarray}
Z_{{\sf q}(k,k')}&=&({\bf n}_{k'} - {\bf n}_{k})\hat{P}_{k,k'} \:,
\nonumber \\
X_{{\sf q}(k,k')}&=&({\bf a}_{k}^\dagger {\bf a}_{k'} +
               {\bf a}_{k} {\bf a}_{k'}^\dagger)\hat{P}_{k,k'}\:,
\label{bqobs}
\end{eqnarray}
and $Y_{{\sf q}(k,k')}=[Z_{{\sf q}(k,k')},X_{{\sf q}(k,k')}]/2i=-i
Z_{{\sf q}(k,k')}X_{{\sf q}(k,k')}$. These observables obey 
commutation/anti-commutation rules identical to 
(\ref{comm})-(\ref{anticomm}). 
Evolutions generated by the Hamiltonians given in (\ref{bqobs}) can be 
readily constructed from the action of optical phase shifters and beam 
splitters respectively, which allows for an easy implementation of 
arbitrary one qubit ($U(2)$) gates via passive linear optics. 

The final ingredients that are required to make the bosonic qubit
useful include the ability to initialize the qubit in the intended logical 
space $\cH_{{\sf q}(k,k')}$, and the ability to read-out the qubit observables.
In {\sc loqc}, the state preparation of qubit ${\sf q}(k,k')$ can be 
accomplished by using a single-photon source to prepare mode $k$ in 
$|0\rangle_k$ and mode $k'$ in $|1\rangle_{k'}$. To measure the bosonic 
qubit it suffices to use a photo-detector on the mode $k$, which 
destructively determines whether one or more photons were present in the 
mode. Further details are in \cite{klm:2000}.

\subsection{The three bit encoded qubit} 

Our second example comes from quantum error correction using stabilizer
codes. Consider the simplest possible situation, where we wish to
protect one quantum bit against single bit-flip errors by encoding it
into three physical qubits.  A quantum code can be specified by a
two-dimensional subspace of the overall state space $\cS$, $\cC=\mbox{span} 
\{ |0_L\rangle, |1_L\rangle\}\subset \cS$. To protect against single
bit-flip errors, a repetition code $\cC$ can be defined by the encoding
\begin{equation}
(c_0 | 0\rangle + c_1 | 1 \rangle ) \otimes |00 \rangle 
\mapsto c_0 |0_L\rangle + c_1 |1_L \rangle =c_0 |000 \rangle + 
c_1 |111\rangle \:. 
\label{code2}
\end{equation}
It is easily checked that $\cC$ satisfies the 
necessary and sufficient conditions for error-recovery with respect to 
the error set 
${\sf E}=\{ E_0=\openone,E_1=X_1,E_2=X_2,E_3=X_3 \}$~\cite{preskill,kl}. 
Note that $E_a=E^\dagger_a=E_a^{-1}$ for errors in ${\sf E}$.
Let $\cV^i$ denote the subspace 
spanned by ${\sf E}|i_L\rangle$, for $i=0,1$, and let us choose as 
orthonormal bases in the $\cV^i$'s the result of applying the
errors to the
logical states (\ref{code2}), {\it i.e.}
\begin{eqnarray}
\cV^0 &=& \mbox{span}\{ |000 \rangle, |100 \rangle, |010 \rangle, 
|001 \rangle \} =  \mbox{span} \{ |v_a^0 \rangle \} \:, \nonumber \\
\cV^1 &=& \mbox{span}\{ |111 \rangle, |011 \rangle, |101 \rangle, 
|110 \rangle \} =  \mbox{span} \{ |v_a^1 \rangle \} \:, 
\label{vi}
\end{eqnarray}
with $ |v_a^i \rangle =E_a |i_L\rangle$, $i=0,1$, $a=0,\ldots,3$. Then
$\cS \simeq \cV^0 \oplus \cV^1$ and a recovery super-operator can be 
explicitly constructed by defining, for each error $E_a \in {\sf E}$,
\begin{equation}
R_a=E_a \sum_{i=0,1}\,  |v_a^i \rangle  \langle v_a^i |\:. 
\label{recovery}
\end{equation}
The fact that the quantum operation $\cR =\{ R_a\}$ defined by
$\cR:\rho\mapsto\sum_aR_a\rho R_a^\dagger$ for density operators
$\rho$ actually restores the state of the encoded qubit after an error
in ${\sf E}$ happens is due to the property that every combination
$R_aX_b$ is a multiple of the identity operation on $\cC$.  Thus, the
basic idea for using this code is to apply $\cR$ after the errors happened.  
While this procedure successfully protects our bit of quantum information, 
it is conceptually dissatisfying, because it would appear that {\sl after} 
the errors happened, but {\sl before} application of $\cR$, the information 
is corrupted by noise.
Is there a representation that clearly separates errors and information 
in such a way that it is clear that the quantum information is never 
affected? In other words, where does the protected qubit reside both before 
and after errors occurred?

The basic insight is to regard the error-correcting code as an appropriate 
{\sl subsystem}~\cite{klv}. This is possible by establishing the mapping
\begin{equation}
|v_a^i \rangle \simeq |i \,\rangle_{\sf q} \otimes   
|v_a^0 \rangle_{\cE} \:, 
\label{qmap}
\end{equation}
for the basis vectors of the $\cV^i$'s (hence $\cS$) introduced before. 
In the right hand-side of (\ref{qmap}), the  $|i \,\rangle_{\sf q} $ 
vectors, $i=0,1$, are taken as basis states of a two-dimensional complex 
space that will serve as the protected qubit state space, while the 
$|v_a^0 \rangle_{\cE}$ store the bit string that uniquely identifies the 
error syndrome. Essentially,
the state $|v_a^0 \rangle_{\cE}$ is meant to fully encode the effect of 
the noise on the code. The correspondence (\ref{qmap}) is a
prescription for decomposing the physical coding space $\cS$ into the 
tensor product of a qubit space $\cQ$ and a syndrome space $\cE$:
\begin{equation}
\cS \simeq \cQ \otimes \cE \simeq \cmplxs^2 \otimes \cmplxs^4 \:. 
\label{code3}
\end{equation}
If we choose to represent the syndrome corresponding to $E_0$ (no
error) by $|00\rangle_{\cE}$, then clearly $\cC \simeq \cQ \otimes
|00\rangle_{\cE}$.  In the representation (\ref{code3}), it is
straightforward to visualize the errors' effect on the code. For an
arbitrary encoded state $|\psi \rangle =c_0 |0_L \rangle +c_1 |1_L
\rangle \in \cC$, we find
\begin{eqnarray}
E_a |\psi \rangle & = &
c_0 |v_a^0 \rangle + c_1 |v_a^1 \rangle
  \simeq  c_0 |0 \rangle_{\sf q} \otimes  |v_a^0 \rangle_{\cE} +   
       c_1 |1 \rangle_{\sf q} \otimes  |v_a^0 \rangle_{\cE} \nonumber \\
 & = & [c_0 |0 \rangle_{\sf q}  +c_1  |1 \rangle_{\sf q}]  
\otimes  |v_a^0 \rangle_{\cE} =  |\psi \rangle_{\cQ} \otimes   
|v_a^0 \rangle_{\cE} \:, \label{errorb}
\end{eqnarray}
where in the last equality the mapping (\ref{code3}) is made explicit. 
By construction, the vector $|v_a^0 \rangle_{\cE}$ depends on $E_a$ alone.
Thus, information in $\cQ$ is completely unaffected by errors in ${\sf E}$: 
the factor $\cQ$ in (\ref{code3}) is the qubit subsystem where the 
protected quantum information resides.

It is worth comparing this picture with the familiar description 
of the error-correcting code based on the stabilizer formalism 
\cite{stabilizer}. In the stabilizer language, the code $\cC$ is 
characterized by its stabilizer group,
\begin{equation}
{\sf S} = \{ \openone, Z_1Z_2,  Z_2Z_3,  Z_1Z_3 \} \:. 
\label{stab}
\end{equation}
{\sf S} is generated by $M_1=Z_1Z_2$ and $M_2= Z_2Z_3$, and the 
code $\cC$ is a joint eigenspace of the two stabilizer generators.
The commutation pattern between each error $E_a$ and the stabilizer 
generators $M_j$ diagnoses the error syndrome completely. 
In our case, it simply reads $E_0 \rightarrow 00$, $E_1
\rightarrow 10$, $E_2 \rightarrow 11$, $E_3 \rightarrow 01$, where
$0,1$ encodes whether the error commutes or anti-commutes with the
corresponding generator, respectively. 

A basis of the state space of the three qubits can be built from joint
eigenvectors of a sufficiently large (maximal) set of commuting operators. 
Such a set can be obtained by adding to $M_1$ and $M_2$ the operator 
$L=Z_1$ (for example). The corresponding joint eigenvectors may be labeled 
$|l,m_1,m_2\rangle$, where, for convenience, we denote by 
$l,m_1,m_2$ not the $\pm 1$ eigenvalues of the corresponding operators 
$L,M_1,M_2$, but the eigenvalues of the projectors onto the $-1$ subspace
{\it e.g.}, the label $l=0$ ($l=1$) corresponds to a $+1$ ($-1$)
eigenvalue of the operator $L$.  The code $\cC$ is then spanned by the 
vectors $|0,0,0\rangle$ and $| 1,0,0\rangle$. The quantum
number $l\in \{ 0,1\}$ can be thought of as labeling a two-state degree 
of freedom. This also translates into a prescription for decomposing the 
overall coding space, 
\begin{equation}
\cS \simeq \cQ' \otimes \cE' \simeq 
\cmplxs^2 \otimes \cmplxs^4 \:, 
\label{code4}
\end{equation}
via a correspondence of the form 
\begin{equation}
|l,m_1,m_2 \rangle \simeq |l\, \rangle_{\sf q'} \otimes |m_1,m_2
 \rangle_{\cE'} \:.
\label{qmap2}
\end{equation}
Note that this prescription is not unique, as it depends on the choice
of phase of the different eigenvectors. (It can be made unique by
considering $L$ and the global flip $X_1X_2X_3$ as generators for the 
algebra of observables for the degree of freedom carried by ${\sf q'}$.)  
The difference between the representations (\ref{qmap2}) and (\ref{qmap}) 
accounts for the different views of error-correcting codes that comes from 
the subsystem idea.  The difference can be explicitly appreciated by 
looking at the correspondences between the states in the two factorizations. 
We have, for instance,
\begin{eqnarray}
| v^0_0 \rangle = |000 \rangle \simeq  
|0 \rangle_{\sf q} \otimes |00\rangle_{\cE} &\simeq& |0\rangle_{\sf q'} 
\otimes |00 \rangle_{\cE'} \nonumber \\
|v^1_0 \rangle = |111 \rangle \simeq 
|1 \rangle_{\sf q} \otimes |00\rangle_{\cE} &\simeq& |1\rangle_{\sf q'} 
\otimes |00 \rangle_{\cE'}\:,
\end{eqnarray}
but
\begin{eqnarray}
|v^1_1 \rangle = |011 \rangle \simeq  
|1\rangle_{\sf q}\otimes |10 \rangle_{\cE} &\simeq& |0\rangle_{\sf q'} 
\otimes |10 \rangle_{\cE'} \nonumber \\
|v^0_1 \rangle = |100 \rangle \simeq  
|0\rangle_{\sf q}\otimes |10\rangle_{\cE} &\simeq& |1\rangle_{\sf q'} 
\otimes |10 \rangle_{\cE'}\:,
\end{eqnarray}
meaning that the degree of freedom ${\sf q'}$ has its values flipped by 
some error combination and is therefore not explicitly protected. 
Specifically, $E_1 [|c_0 |0_L \rangle + c_1 |1_L \rangle] \simeq 
[|c_0 |1 \rangle_{\sf q'} + c_1 |0 \rangle_{\sf q'}]\otimes 
|10 \rangle_{\cE'}$. The error can be undone by a flip of ${\sf q'}$ 
conditional on the state of $\cE'$. Thus, while in this picture error 
detection and error recovery necessarily appear as two separate steps, 
the earlier factorization (\ref{qmap}) makes sure that the need for an 
explicit error correction step is removed once information is properly 
stored: to first order in time, the effect of the noise is only to  
``heat'' the ancillary degrees of freedom carried by the syndrome 
subsystem $\cE$, and all that is required to recover from errors is 
to ``reset'' the latter subsystem to the state $|00\rangle_{\cE}$. 
This is exactly what the recovery quantum operation $\cR$ accomplishes, 
as is clear by representing every operator $R_a$ in (\ref{recovery}) 
using the mapping (\ref{qmap}), {\it i.e.}
\begin{equation}
R_a \simeq \openone^{(\cQ)} \otimes |v^0_0 \rangle 
\langle v^0_a|^{(\cE)} \:.
\label{recovery2}
\end{equation}
		
The best method for characterizing the protected qubit living in $\cQ$
is in terms of its observables. This is easily done by starting with
the obvious observables restricted to the code subspace:
$Z_{\cC}=|0_L\rangle \langle 0_L| -
|1_L\rangle \langle 1_L|$, $X_{\cC}=|0_L\rangle \langle 1_L| +
|1_L\rangle \langle 0_L|$. The $Z$ and $X$ observables for the subsystem 
associated with ${\cQ}$ can then be obtained by applying the errors in 
the following way:
\begin{eqnarray}
Z_{\sf q} & = & Z_{\cC} +E_1 Z_{\cC} E_1 +E_2 Z_{\cC} E_2 + 
E_3 Z_{\cC} E_3 \:, \nonumber \\
X_{\sf q} & = & X_{\cC} +E_1 X_{\cC} E_1 +E_2 X_{\cC} E_2 + 
E_3 X_{\cC} E_3 \:. 
\label{obscode}
\end{eqnarray}
Note that the required behavior of $Z_{\sf q}$, $X_{\sf q}$ on the 
subsystem $\cQ$ follows from the property that each operator  
$E_a {\cO}_{\cC} E_a$ in (\ref{obscode}), 
${\cO}_{\cC}= Z_{\cC}$ or $X_{\cC}$, has the intended action on 
states of the form $E_a |i_L \rangle$, $i=0,1$. 
The relevant commutation and anti-commutation rules
(\ref{comm})-(\ref{anticomm}) are readily verified. In addition to
belonging to the set of operators commuting with the stabilizer group 
(the so-called normalizer ${\sf N(S)}$ \cite{stabilizer}), the above qubit 
observables have the property that they also commute with every operator 
that is a repeated combination of a reset operator followed by an error, 
{\it e.g.} $\ldots E_{b'} R_{a'} E_b R_a$. This sheds light on the algebraic
nature of our three-bit subsystem, which is characterized as a
{\sl noiseless subsystem} of the multiplicative algebra $\cA$ constructed
from recovery operators followed by errors\footnote{The multiplicative (or associative) algebra $\cJ$ 
generated by a linear set of operators $\cJ_1=\mbox{span}\{ 
\openone,J_1,J_2,\ldots \}$ contains all the linear 
complex combinations of products of operators in $\cJ_1$.}.
The fact that {\sl every} quantum error-correcting code can be pictured as 
a noiseless subsystem of a suitable operator algebra is established 
in~\cite{klv}.

\subsection{The three spin-$1/2$ symmetric qubit}

Consider a system $S$ composed by three physical, distinguishable spin
$1/2$ particles. Suppose that the interaction with the environment $B$
results in a particularly simple form of noise, {\sl collective}
noise, where $B$ couples in a symmetric way to each
spin~\cite{zanardi:1997}. A relevant example of this situation occurs 
when the spins couple identically to a fluctuating magnetic field.
Our task is to devise a scheme for embedding in $S$ a qubit that is 
protected to noise. 

For collective noise, the interaction Hamiltonian $H_{SB}$ can we
written as
\begin{equation}
H_{SB}=\sum_{\alpha={\sf x}, {\sf y}, {\sf z}} \, S_\alpha \otimes
B_\alpha \:,
\label{hsb}
\end{equation}
where $S_\alpha=\sum_i \sigma^{(i)}_\alpha/2$, $\alpha={\sf x}, {\sf y},
{\sf z}$, are the components of the total spin angular momentum 
and the $B_\alpha$'s are linearly independent environment operators. Note
that the operators $S_\alpha$ are the generators of a Lie group $SU(2)$ 
which corresponds to global spatial rotations of the spins, and can be 
identified with the familiar $SU(2)$ of angular momentum theory 
\cite{cornwell}. Suppose that the self-Hamiltonian of the spins can also 
be expressed in terms of the $\{S_\alpha \}$.
One possibility for constructing a protected qubit is to look for a pair 
of simultaneous degenerate eigenstates of the $S_\alpha$'s 
\cite{zanardi:1997,lidar} {\it i.e.}, 
\begin{equation}
S_\alpha | \psi_l \rangle =c_\alpha  |\psi_l \rangle\:,
\hspace{5mm} \alpha={\sf x,y,z}\,,\: l=1,2\:. 
\label{deg}
\end{equation}
Since the eigenvalues $c_\alpha$ do not depend on $l$, the two
eigenstates cannot be distinguished by the environment. Thus, if they
can be found, they define a basis of a protected qubit's state space.
In our case, one finds that (\ref{deg}) can be only satisfied with
$c_\alpha=0$\footnote{Technically, this follows from the fact
that the operators $S_\alpha$ span a semi-simple Lie algebra $su(2)$
\cite{cornwell}.}, meaning that the states $|\psi_l \rangle $ belong
to the so-called $S^2=0$ singlet representation of $su(2)$ {\it i.e.}
they are invariant under $SU(2)$ rotations.  Unfortunately, no state
of three spin $1/2$ particles obeys this invariance condition, and a
minimum number of four physical spin $1/2$ particles is required for
the singlet representation to occur with degeneracy at least
two~\cite{zanardi:1997}. In spite of this impossibility to find a {\sl
subspace} of the three-spin state space $\cS$ which is immune against
noise, it turns out that we are still able to construct a {\sl
noiseless subsystem} of $\cS$~\cite{klv,vkl,zanardi2,defilippo} by
considering observables commuting with the $S_\alpha$'s.  The idea,
which we describe next, is to identify in $\cS$ a protected degree of
freedom.

Since the Hamiltonians $S_\alpha$ generate the spatial rotation group 
acting symmetrically on the three spins, it is natural to decompose the 
state space $\cS$ of the spins according to the total angular momentum 
$S^2=\sum_\alpha S^2_\alpha$:
\begin{equation}
\cS = \cH_{3/2} \oplus \cH_{1/2} \:, 
\label{cg}
\end{equation}
where the subspaces $\cH_S$ are the eigenspaces of $S^2$ corresponding
to angular momentum $S=3/2,1/2$, with $S^2=S(S+1)$.  They have
dimension dim$\,(\cH_{3/2})=(2\cdot 3/2+1)=4$,
dim$\,(\cH_{1/2})=2(2\cdot 1/2+1)=4$, respectively.  Let us focus on
the $S=1/2$ component.  The fact that $\cH_{1/2}$ has dimension four
implies that the two-dimensional spin-$1/2$ representation of the
rotation group occurs twice, meaning that physically there are two
distinct, equivalent routes for generating angular momentum
$S=1/2$\footnote{In representation-theoretical terms, the appropriate
decomposition of the state space is obtained as the Clebsch-Gordan sum
of irreducible representations of $su(2)$ which, for our three-spin
system, reads $\cS \simeq \cD_{3/2} \oplus \cD_{1/2}\oplus \cD_{1/2}$
\cite{cornwell}.}.  A basis of states for $\cH_{1/2}$ is constructed
by considering joint $S^2$, $S_z$-eigenvectors, $\{ |\lambda, s_z
\rangle_{1/2} \: | \: \lambda=0,1;\: s_z=\pm 1/2 \}$, where $s_z$ is
the $S_z$-eigenvalue and the quantum number $\lambda$ identifies which
of the distinct routes the corresponding eigenvector belongs
to. Because angular momentum operators $S_\alpha$ are confined to act
non-trivially and equivalently within each route, the $S_\alpha$ have
an identical, diagonal action on the degree of freedom supported by
the quantum number $\lambda$. This degenerate behavior of noise
operators with respect to $\lambda$ is exactly the two-fold degeneracy
we are looking for.

A better grasp of the protected structure which is emerging in $\cH_{1/2}$ 
is obtained by establishing the mapping 
\begin{equation}
|\lambda, s_z \rangle_{1/2} \simeq |\lambda \rangle_{\sf q} \otimes
| s_z \rangle_{1/2} \:, 
\label{map}
\end{equation}
where now $\{ |\lambda \rangle_{\sf q} \}$ is an orthonormal basis in
$\cmplxs^2$ and $| s_z \rangle_{1/2}$ is the $S_z$-eigenvector with
eigenvalue $s_z$. Under such an identification, the subspace $\cH_{1/2}$ 
can be represented as 
\begin{equation}
\cH_{1/2}  \simeq \cH_{\sf q} \otimes \cD_{1/2} \simeq \cmplxs^2
\otimes \cmplxs^2\:. 
\label{factors} 
\end{equation}
The actions of the noise operators $S_\alpha$ on $\cH_{1/2}$ take a 
correspondingly simple form,
\begin{equation}  
S_\alpha   \simeq \openone^{(\cH_{\sf q})} \otimes 
\sigma(\alpha) \:,
\label{actions}
\end{equation}
where $\sigma(\alpha)$ is a unit linear combination of Pauli operators
which depends on the choice of basis states in $\cH_{\sf q}$. 
These algebraic identities provide the starting point for identifying 
$\cH_{1/2}$ as the state space of an effective bi-partite system, and 
for associating an abstract subsystem to each factor in the tensor 
product (\ref{factors}). In particular, by virtue of the identity action 
of noise operators in (\ref{actions}), the left factor $\cH_{\sf q}$ 
provides a {\sl noiseless} subsystem where a qubit can safely reside, 
protected from noise for (ideally) arbitrarily long times.   

Because the $\cmplxs^2$-basis vectors $|\lambda\rangle_{\sf q} $ are 
left arbitrary in (\ref{factors}), various realizations are possible as 
basis states of our three-spin noiseless qubit. Two convenient choices 
are listed below:
\begin{eqnarray}
|\tilde{0} 
\rangle_{\sf q} \otimes |+1/2\rangle_{1/2} & = & {1 \over \sqrt{2}}\,
(|01\rangle- |10\rangle) |0 \rangle  \:, \nonumber \\
|\tilde{1} 
\rangle_{\sf q} \otimes |+1/2\rangle_{1/2} & = & 
{1 \over \sqrt{6}}\,
(2|001\rangle - |010\rangle - |100 \rangle )\:,  \nonumber \\
|\tilde{0} 
\rangle_{\sf q} \otimes |-1/2\rangle_{1/2} & = & {1 \over \sqrt{2}}\,
(|10\rangle- |01\rangle) | 1\rangle  \:, \nonumber \\
|\tilde{1} 
\rangle_{\sf q} \otimes |-1/2\rangle_{1/2} & = & {1 \over \sqrt{6}}\,
(2|110\rangle - |101\rangle - |011 \rangle )\:,  
\label{basis1}
\end{eqnarray}
which can be easily built up from singlet/triplet states of spins 1 and 2
\cite{kempe}, or
\begin{eqnarray}
|0 \rangle_{\sf q} \otimes |+1/2\rangle_{1/2} & = & {1 \over \sqrt{3}}\,
(|001\rangle + \omega |010 \rangle + \omega^2 |100 \rangle) \:, \nonumber\\
|1 \rangle_{\sf q} \otimes |+1/2\rangle_{1/2} & = & {1 \over \sqrt{3}}\,
(|001\rangle + \omega^2 |010 \rangle + \omega |100 \rangle) \:,  \nonumber \\
|0 \rangle_{\sf q} \otimes |-1/2\rangle_{1/2} & = & {1 \over \sqrt{3}}\,
(|110\rangle + \omega |101 \rangle + \omega^2 |011 \rangle) \:,  \nonumber \\
|1 \rangle_{\sf q} \otimes |-1/2\rangle_{1/2} & = & {1 \over \sqrt{3}}\,
(|110\rangle + \omega^2 |101 \rangle + \omega |011 \rangle) \:,
\label{basis2}
\end{eqnarray}
with $\omega=e^{2\pi i /3}$, which connects directly with standard
basis states for the two-dimensional irreducible representation {\bf
D}$^{(1)}$ of the permutation group ${\sf S}_3$ acting on the
spins~\cite{peres}. In (\ref{basis1})-\ref{basis2}), the notations 
$\{ |\tilde{0} \rangle_{\sf q}, |\tilde{1} \rangle_{\sf q} \}$ and
$\{ |0 \rangle_{\sf q}, |1 \rangle_{\sf q} \}$ have been used to 
account for the different choices of the basis states 
$\{ |\lambda\rangle_{\sf q} \}$ in $\cH_{\sf q}$. It is important to 
realize that a possibly mixed state of $\cS$ of the form 
$\rho=|\psi\rangle \langle \psi| \otimes \rho_{1/2}$, where 
$|\psi\rangle =c_0 |0 \rangle_{\sf q} + c_1 |1 \rangle_{\sf q}$ 
and $\rho_{1/2}$ is an arbitrary density operator on $\cD_{1/2}$, 
is a pure state of the qubit living in $\cH_{\sf q}$.

The bases given above establish an equivalence between the state 
spaces of a pair of two-state systems and the subspace $\cH_{1/2}$ of the three
spins' state space $\cS$. The method can be systematized and the
introduction of a different state space can be avoided by directly
considering the set of physical observables available for the three spins.
Observables that are not affected by the interaction operators
$S_\alpha$ are the scalars under spatial rotations, which are given by
\begin{eqnarray}
s_{12} &=& \vec{\sigma}^{(1)} \cdot \vec{\sigma}^{(2)} 
= X_1X_2 + Y_1Y_2 +Z_1Z_2 \:, \nonumber \\
s_{23} &=& \vec{\sigma}^{(2)} \cdot \vec{\sigma}^{(3)} 
= X_2X_3 + Y_2Y_3 +Z_2Z_3 \:, \nonumber \\
s_{31} &=& \vec{\sigma}^{(3)} \cdot \vec{\sigma}^{(1)} 
= X_3X_1 + Y_3Y_1 +Z_3Z_1 \:. \label{scalars} 
\end{eqnarray} 
Since these scalars commute with the interactions, their expectations
are protected from noise. If we can combine them so that their
expectations define states of a qubit, a noiseless qubit has been found.
It is sufficient to form combinations that algebraically behave like 
the Pauli operators. Let $P_{\sf q}: \cS \mapsto \cH_{\sf q}$ denote the 
projector over $\cH_{\sf q}$,
\begin{equation}
P_{\sf q} = {\openone \over 2} -{1\over 6} \,
(s_{12}+s_{23}+s_{31} ) \:,
\label{pq}
\end{equation}
$\openone$ denoting the identity operator over $\cS$ and $P_{\sf q}$ 
satisfying $P_{\sf q}=_{\sf q} \openone_{\sf q}$ when restricted to 
$\cH_{\sf q}$. Then the following choice of observables works and 
corresponds to the $\omega$-basis introduced above:
\begin{eqnarray}
X_{\sf q} & = & {1\over 6} \,(2s_{12}- s_{23}- s_{31}) P_{\sf q} = 
 E_{12}P_{\sf q} \:, \nonumber \\
Y_{\sf q} & = & -{\sqrt{3}\over 6} \, (s_{23}- s_{31}) P_{\sf q} \:,
\label{obs3}
\end{eqnarray}
where $E_{12}=(\openone + \vec{\sigma}^{(1)} \cdot \vec{\sigma}^{(2)})/2$ 
is the unitary operator swapping the first two spins.  $Z_{\sf q}$ is given
explicitly as $Z_{\sf q}= [X_{\sf q},Y_{\sf q}]/2i= \sqrt{3}/6\,
\tau_{123}$, $ \tau_{123}$ being the totally antisymmetric operator
$\tau_{123}=\sum_{\alpha\beta\gamma} \varepsilon_{\alpha\beta\gamma}
\sigma^{(1)}_\alpha \sigma^{(2)}_\beta \sigma^{(3)}_\gamma$,
$\alpha,\beta,\gamma\in \{ {\sf x,y,z}\}$.  Similar expressions hold
for the singlet-triplet basis (\ref{basis1}), {\it e.g.}  
$X_{\sf q}= {\sqrt{3}} \, (s_{23}- s_{31}) P_{\sf q} /6$, 
and $Z_{\sf q}=- E_{12} P_{\sf q}$. Since the commutation and 
anti-commutation rules obeyed by the observables are identical to 
(\ref{comm})-(\ref{anticomm}), the operator algebra they generate by 
multiplication and Hermitian conjugation is the algebra of a two-state 
quantum system.

An important simplification to the above discussion is worth
mentioning.  This occurs when not only $S^2$ but also $S_z$ are good
({\it i.e.} conserved) quantum numbers for the dynamics of the three
physical spins. In this case, using appropriate encoded states rather 
than the physical three-spin states is motivated by the possibility
of achieving universal control by means of purely angular momentum 
preserving quantum gates. Note that three is the minimal number of 
physical qubits allowing for simultaneous eigenvectors of $S^2,S_z$ 
with degeneracy at least equal to two. The resulting qubits are the
basic ingredients in a recent proposal for universal quantum
computation via exchange interactions \cite{kempe}.
The mathematical description of these qubits is simpler than the one 
of the noiseless three-spin case, as the full subsystem structure 
(\ref{factors}) is not required. The relevant observation is that if no 
mixing between the sets of states in $\cH_{1/2}$ corresponding to
$s_z=\pm 1/2$ occurs, a further decomposition applies within the 
$\cH_{1/2}$ subspace:
\begin{equation}
\cH_{1/2} \simeq  \cmplxs^2 \otimes \cD_{1/2} \simeq  \cmplxs^2 
\otimes \Big(\cD_{1/2}^{(+)} \oplus \cD_{1/2}^{(-)}\Big) \simeq \cmplxs^2 
\otimes (\cmplxs \oplus \cmplxs) \simeq \cmplxs^2 \oplus \cmplxs^2\:.
\label{summands}
\end{equation}
This means that the two subspaces spanned by $\{ |\lambda, +1/2
\rangle_{1/2} \}$ and $\{ |\lambda, -1/2 \rangle_{1/2} \}$,
$\lambda=0,1$ (the first and second pair of states in either
(\ref{basis1}) or (\ref{basis2})) are {\sl individually} capable of
encoding a qubit. Of course, none of these qubits will retain the
robust behavior against all collective noise. Over each
$\cmplxs^2$-summand of (\ref{summands}), the relevant qubit
observables are still given by the combinations of scalars determined
above. Note that, once a fixed angular momentum component is chosen,
say $s_z=+1/2$, states belonging to $\cmplxs^2\otimes \cD_{1/2}^{(-)}$
as well as to $\cH_{3/2}$ behave as ``non-qubit'' levels similarly to
what we encountered in the case of the bosonic qubit.  In particular,
this results in a many-qubit state space structure analogous to
(\ref{mqss}), with a non-trivial summand $\tilde{\cH}$ accounting for
all the possible ``non-qubit'' configurations.

Explicit prescriptions are mentioned in \cite{kempe} for both
initializing the logical qubit to the $|\tilde{0}\rangle_{\sf q}$
state, and for implementing the final qubit measurement. Initialization
may be achieved by relying on the natural equilibration processes of
the physical spins in the presence of a polarizing magnetic field,
while measurement of the qubit in the $\{|\tilde{0}\rangle_{\sf q},
|\tilde{1}\rangle_{\sf q}\}$ basis ($s_z=+1/2$) may be accomplished by
determining the singlet vs. triplet state of spins 1, 2.

\section{Criteria for a qubit}
\label{sect:criteria}

So what is a qubit? From the mathematical point of view, the main
lesson that emerges from the above examples, and from the many others
that can be analyzed along similar lines, is that the algebraic
notion of a {\sl subsystem} provides the most general framework for
capturing the variety of ways in which qubits may be constructed.
Subsystems, intended as factors (in the tensor product sense) of
subspaces of a possibly larger state space, are the structures 
where we have identified qubits throughout our analysis. 
Because the information-carrying qubits can be far removed
from the ``natural'' physical systems that a device is based on,
states of qubit subsystems often look complicated when expressed
in a basis associated with the physical degrees of freedom, making it
difficult to recognize their properties and dynamical behavior.
The situation is much simpler, and the description more compact, 
if the qubit is realized in an {\sl operator} sense through its 
{\sl observables}.

The idea of describing quantum systems in terms of operators forming a
complex associative algebra, whose Hermitian elements provide the
system's observables, underlies the operator approach to quantum and
quantum-statistical mechanics \cite{thirring}.  Similarly, the general
definition of a subsystem is motivated \cite{klv} by a fundamental
representation theorem for finite dimensional associative operator 
algebras closed under Hermitian conjugation, stating that for any such 
algebra $\cA$ a direct sum representation of the overall state space 
$\cS$ exists,
\begin{equation}
\cS \simeq  \bigoplus_i{\cC_i \otimes \cD_i} \:,
\label{iso}
\end{equation}
in such a way that $\cA$ has identity action over each of the factors 
$\cC_i$'s, $\cA \simeq \oplus_i \openone^{(\cC_i)} \otimes \mbox{End}
(\cD_i)$. Thus, if $\cA$ is the algebra constructed from noise operators
({\sl interaction algebra} \cite{klv}), a noiseless factor (subsystem) 
$\cC_i$ is naturally characterized by an irreducible representation of the 
so-called {\sl commutant} $\cA'$, which is formed from the operators 
commuting with everything in $\cA$. 

Motivated by this general perspective, a necessary condition for
having a qubit is that it is a subsystem whose associative
operator algebra is identical with (isomorphic to) the ``right''
operator algebra of a two-state quantum system, {\it i.e.}  one whose
generators satisfy the set of composition rules specified in
(\ref{comm})-(\ref{anticomm}). Notice that this requirement is
stronger than the one based on the $su(2)$ commutation rules
(\ref{comm}) alone.  While the latter are crucial in determining the
appropriate {\sl Lie}-algebraic structure (thereby obtaining the
correct symmetry properties) of our qubit, (\ref{comm}) and
(\ref{anticomm}) together are necessary (and sufficient) for ensuring
the correct {\sl associative} structure of the relevant algebra.  In
particular, the condition $\cO_{\sf q}^2=\openone_{\sf q}$ should hold
for the generating observables $\cO_{\sf q}$ to ensure that the
appropriate representation in terms of an abstract spin-1/2 particle
is realized.

How sensible is this definition from a physical standpoint? Looking
back at our examples once more, one observes that identifying the
correct algebraic structure does not guarantee by itself the ability of
using the associated subsystem as a qubit. In a sense, this is only
the first step toward constructing a qubit. Suppose, however, that we 
did actually succeed at determining a set of observables which generate
the correct operator algebra, and suppose, in addition, that we have
the capabilities for implementing the following manipulations:
\begin{itemize}
\item {\sl Unitary control}$-$ Apply the observables as Hamiltonians to
effect universal control operations on the subsystem.

\item {\sl Initialization}$-$ Apply suitable non-unitary control, {\it
i.e.}  a quantum operation, so as to leave the subsystem in a state
whose expectation on the observables matches that of $|0\rangle$ for
some choice of the observable $|0\rangle \langle 0|$.

\item {\sl Read-out}$-$ Strong version: Perform von Neumann projective
measurements of the subsystem observables, which together with unitary
control implies the ability to initialize. Weak version: Perform weak
ensemble measurements of the subsystem observables.
\end{itemize}
Then what we have constructed is, for all practical purposes, a qubit.

\section{Summary and Conclusions}

We have provided an operational guideline for constructing qubits 
in physical systems. Our analysis emphasizes the role of operator
algebras and observables as the most powerful and comprehensive language 
to be used for defining qubits in full generality. One obvious 
implication is that a qubit ends up being a much more versatile and 
general object than one might at first conceive of. In a broader context, 
it was recently argued by Steane~\cite{steane:2000} that a picture in 
terms of operators rather than state vectors could provide a more insightful 
perspective for understanding various aspects of quantum information 
processing, including the origin of the apparent improvements in efficiency 
over classical information processing. In a sense, a definition of the
qubit in general operator terms can then be regarded as the first 
necessary step of this program.

\ack
This work was supported by the NSA and by the DOE under contract 
W-7405-ENG-36.

\end{document}